\documentclass[12pt]{article} 
\usepackage{epsfig}
\newlength{\dinwidth}
\newlength{\dinmargin}
\newcommand{\ba}{\begin{array}}
\newcommand{\ea}{\end{array}}
\newcommand{\be}{\begin{equation}}
\newcommand{\ee}{\end{equation}}

\def\lsim{\mathrel{\rlap{\lower4pt\hbox{\hskip1pt$\sim$}}
    \raise1pt\hbox{$<$}}}         
\def\gsim{\mathrel{\rlap{\lower4pt\hbox{\hskip1pt$\sim$}}
    \raise1pt\hbox{$>$}}}         

\def\Pom{{\bf I\!P}}

\begin{document}
\vspace*{1cm}
\begin{center}  \begin{Large} \begin{bf}
Diffraction in DIS on nuclear targets\\
  \end{bf}  \end{Large}
  \vspace*{5mm}
  \begin{large}
N.N.Nikolaev$^{a,b,c}$, W.Sch\"afer$^{b}$, B.G.Zakharov$^{c}$,
V.R.Zoller$^{d}$\\
  \end{large}
  \vspace*{3mm}
$^a$ ITKP, Universit\"at Bonn, Nu{\ss}allee 14-16, Bonn D-53115, FRG
\\
$^b$ Inst. f. Kernphysik, KFA J\"ulich, D-52425 J\"ulich, FRG\\
$^c$ L.D.Landau Inst. Theor. Physics, Kosygina 2, 117334 Moscow, Russia\\
$^d$ Inst. Theor. exp. Physics, B.Cheremushkinskaya 25, 117259 Moscow, Russia\\
\end{center}
\begin{quotation}
\noindent
{\bf Abstract:}
We discuss the implications of a strong enhancement of diffraction
for multiproduction in DIS off nuclear targets. The predicted
effects are large and observable at HERA. We present the prediction 
of a large tensor structure function $b_{2}(x)$ of the deuteron 
which does not vanish at small $x$.  
 
\end{quotation}
%
\noindent
{ \large \bf Diffractive DIS on nuclei is a huge effect}

At the DESY Mini School in May 1994 one of the present authors 
(NNN) presented a prediction of a strong nuclear enhancement 
of diffractive DIS (DDIS), which on a strongly absorptive nuclei 
must reach $f_{D} \sim 50\%$ of the total DIS rate \cite{NNNDESY}! 
The participants may remember how it has been ridiculed by fellow 
experimentalists as an utterly useless prediction and at 
the Round Table discussion some of the fellow theorists 
stated it must be wrong. It took slightly more than
a year for a serious discussion on the electron-nucleus 
option at HERA to be on the floor.

The detailed evaluation of nuclear enhancement of 
DDIS is published in \cite{NZZgap}. The argument for the 
enhancement goes as follows: The microscopic QCD mechanism 
of DDIS is a grazing, quasielastic scattering of multiparton Fock 
states of the photon on the target proton, which is best described
viewing these Fock states as systems of color dipoles spanned 
between quarks, antiquarks and gluons \cite{NZ91}.
The crucial finding is that for transverse photons DDIS is 
dominated by the contribution from soft dipoles 
$r\sim 1/m_{f}$, still the  $1/Q^{2}$ leading twist behavior of 
$\sigma_{T}^{D}$ is a rigorous QCD result \cite{NZ91}.
The rest of the story is simple: DDIS on nuclei essentially
amounts to an elastic (coherent DDIS) and small admixture 
of quasielastic (incoherent DDIS) scattering of soft dipoles
off a target nucleus. For soft dipoles and strongly absorbing 
targets
\be
\sigma^{D}\sim \sigma^{ND} \sim {1\over 2}\sigma^{tot}\, .
\label{eq:1}
\ee
Here ND stand for the non-diffractive and/or non-LRG DIS. In 
the conventional $A^{\alpha}$ parameterization of nuclear 
cross sections, we find $\alpha \approx 0.9$ for the total DIS
at moderately small $x\sim 10^{-3}$ 
(or $\alpha \approx -0.1$ for the structure function per bound
nucleon) and very large $\alpha^{D}_{coh}\approx$0.25-0.3 for the 
coherent DDIS per bound nucleon. Finally, for the incoherent
DDIS per bound nucleon we predict $\alpha_{inc}^{D} \approx -0.4$.
For the reference, for the carbon target $\sigma_{coh}^{D}:
\sigma_{inc}^{D} \approx 2.2 : 1$.

Coherent DDIS $e A\rightarrow e'XA$ is at work for 
$x,x_{\Pom} \lsim 0.1\cdot A^{-1/3}$ which is precisely
the kinematical range of the $eA$ collisions at HERA.
What are the consequences of this striking nuclear
enhancement of DDIS? Are they observable at HERA?
\\

\noindent 
{ \large \bf DDIS and $A^{\alpha}$ physics in 
multiproduction off nuclei}

The multiproduction off nuclei is usually discussed in terms
of the normalized Feynman $z$ and/or (pseudo)rapidity $\eta$ 
distributions $R(z)= [dn/dz]_{A}/[dn/dz]_{N}$ and
$R(\eta)= [dn/d\eta]_{A}/[dn/\eta]_{N}$. In the hadron-nucleus 
collisions, the more hadronlike is a hadron, i.e., larger is 
$\sigma_{tot}^{hN}$ the stronger is the shadowing in 
$\sigma_{tot}^{hA}$, the higher is the average multiplicity,
the larger is $R(\eta)>1$ at mid-rapidity, the stronger is 
nuclear attenuation of the projectile fragments $R(z\sim 1)
< 1$, the stronger is the hadronic activity in the target nucleus
region (for the review see \cite{NNNUFN}. We predicted a striking 
reversal of this trend when going from pointlike photons in the
nonshadowing (NS) region of $x \gsim 0.1$ to the hadronlike
photons in the shadowing (SH) region of small $x\lsim 0.03-0.01$ 
\cite{NZZgap}. 

Indeed, in coherent DDIS the target nucleus 
remains in the ground state, there is vanishing
hadronic activity in the nucleus region and coherent DDIS falls 
into the LRG category. Incoherent DDIS also contributes to 
LRG events but the fraction of incoherent DDIS rapidly decreases 
with A \cite{NZZgap}.   
Because the DDIS rate increases at small $x$, hadronic activity
in the nucleus region must decrease with the decreasing $x$, 
i.e., for more hadronlike photons. The predicted \cite{NZZgap} 
effect is large, $\sim 30 $\%, and has been fully 
confirmed by the E665 $\mu Xe$ scattering data \cite{E665muXe}.
The very sharp $t$ distribution permits an unambiguous selection
coherent DDIS. HERA experiments can extend these measurements 
to a much smaller $x$ and higher $Q^{2}$ on a broad range of
nuclei. 

\begin{figure}[!htb]
\begin{center}
\epsfig{file=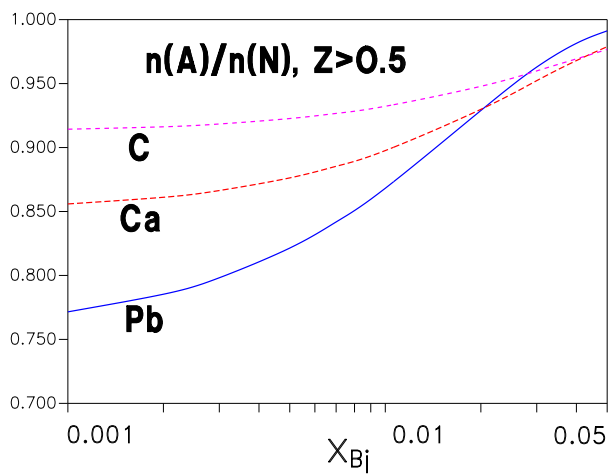,width=8cm}
\vspace{-.5cm}
\end{center}
 \caption[junk]{{\it
Our predictions for attenuation of forward hadrons
in ND DIS off nuclei as a function of $x_{bj}$}.}
\end{figure}

Now consider $R(z)$ for the generic, DDIS/ND unseparated, DIS 
off a nucleus. The very forward hadrons with $z\sim 1$ are predominantly 
the diffraction dissociation products \cite{ZollerDD}. It is 
precisely the part of the phase space usually discussed in 
terms of the fragmentation of the isolated struck quark and the 
Landau-Pomeranchuck-Migdal effect (for the quantum theory 
of the LPM effect see \cite{BGZLPM}). Our point is that DDIS 
completely invalidates this interpretation in the $SH$ region 
of small $x$. Indeed, because of (1) for {\sl heavy, strongly
absorbing} nuclei we predict a {\sl universality}
of the $z$-distributions, $R(z)=1$, which derives from the 
nuclear enhancement of DDIS, the Landau-Pomeranchuck effect is 
irrelevant for this prediction. The corrections to $R(z)=1$ for 
slight nuclear distortions of the mass spectrum in DDIS must be 
marginal with one subtle exception: because of the $M^{2}$
dependent factorization scale \cite{GNZcharm} and related 
effects of color transparency \cite{KNNZ93} DDIS into small masses
$M^{2} \ll Q^{2}$ the vector mesons included will have a much
steeper $A^{\alpha}$ dependence with the exponent
$\alpha \approx $1.35-1.4. Therefore, in the narrow region
of $z\rightarrow 1$ to which only the small mass excitations
can contribute, we predict $R(z)> 1$ !

The ND DIS is the counterpart of inelastic hadron-nucleus 
scattering and our principal prediction is that nuclear effects 
in ND events must be similar to those seen in $\pi A, pA$ collisions: 
$R_{ND}(z) < 1$ at $z\sim 1$ and $R(\eta) > 1$ in the
mid-rapidity region. In particular, hadronic final states 
will be a superposition of subprocesses with the $\nu$-fold
average multiplicity ($\nu$ cut pomerons along the 
lines reviewed in \cite{NNNUFN}. 
Only the $\nu=1$ subprocesses contribute to production of
large-$z$ particles. In Fig.~1 we show the expected
$x$ dependence of nuclear attenuation of forward hadrons
with $z> 0.5$, in the parameterization $R(z) \approx
A^{\alpha(z)}$ for the very forward hadrons and for a sufficiently
small $x \sim 10^{-3}-10^{-4}$ we predict $\alpha(z)
\sim -0.15$. For the central, mid-rapidity region we predict
$R(\eta) \sim A^{0.15}$. The mean multiplicities in the
ND DIS are dominated by central production and must exhibit 
similar $A$ dependence. Such a strong nuclear effects can 
easily be observed at HERA.

Nuclear enhancement of coherent DDIS implies enhancement of low
multiplicities, whereas the enhancement of central production
implies the enhancement of high multiplicities. Consequently,
we predict a substantial broadening of the multiplicity 
distributions, although the overall nuclear increase of the
average multiplicities can be relatively weak.  
\\

\noindent
{\large \bf DDIS and the tensor spin 
structure functions of the deuteron}

For DIS of {\sl unpolarized} leptons off the spin-1 deuterons 
one can define the tensor spin structure function \cite{Jaffe}
$$
b_{2}(x,Q^{2})={1\over 2} \left[F_{2}^{+}(x,Q^{2})+
F_{2}^{-}(x,Q^{2})-2F_{2}^{0}(x,Q^{2})\right]\, ,
$$
where $+,-,0$ refer to the deuteron spin projection on
for instance the $\gamma^{*}D$ collision axis.
One can similarly define the longitudinal tensor structure
function $b_{L}(x,Q^{2})$.

In the usually discussed impulse approximation the tensor parton 
densities vanish for the $S$-wave bound state. The $S-D$ 
interference makes the nucleon momentum distributions and 
folding corrections to structure functions different for
$\pm,0$ polarizations, but at moderately small $x$ the corresponding
tensor asymmetry is a per mill and even smaller effect and
vanishes at $x\rightarrow 0$ \cite{Umnikov}. 

An interesting finding is that an order in magnitude 
larger tensor spin structure
function is generated by DDIS via shadowing corrections \cite{Schaefer}.
Our point is that nuclear shadowing depends on the alignment 
of nucleons in the deuteron, which because of the $S-D$ interference
is different for $\pm$ and 0 polarization states. The striking finding
is that unlike all other spin asymmetries which are well known to vanish 
in the limit $x\rightarrow 0$, the tensor spin asymmetry is finite
and even rises at small $x$. Besides the spin-alignment dependent
shadowing for DDIS, another source of the tensor asymmetry is
DIS  off pions in the deuteron. The number of pions in the deuteron 
also depends on the deuteron spin state. Our predictions for the 
tensor asymmetry are shown in Fig.~2. 
The pion effects have a 
sign opposite to that of the diffractive NS effects and dominate
at $x \gsim 10^{-2}$, diffractive NS takes over at $x\lsim 10^{-2}$.

We predict a nonvanishing $b_{L}(x,Q^{2})$. Typically, the ratio
$R_{LT}=b_{L}/b_{2}\sim 0.2$ and is very similar to the familiar ratio
$\sigma_{L}/\sigma_{T}$ for the nucleon target.

{\bf Acknowledgment.} NNN thanks Prof. U.Mei{\ss}ner for the hospitality
at the Inst. Theor. Kernphysik of the Univ. of Bonn. The work of NNN 
is supported by the DFG grant ME864/13-1.
\newpage
\begin{figure}[!htb]
\begin{center}
\epsfig{file=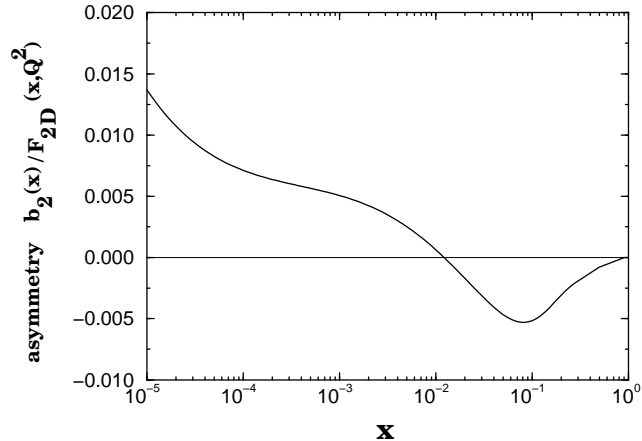,width=8.5cm}
\vspace{-.5cm}
\end{center}
 \caption[junk]{{\it
Our predictions for the tensor asymmetry in DIS of unpolarized
leptons off tensor polarized deuterons ($Q^{2}=10$\,GeV$^{2}$)}.}
\end{figure}

\end{document}